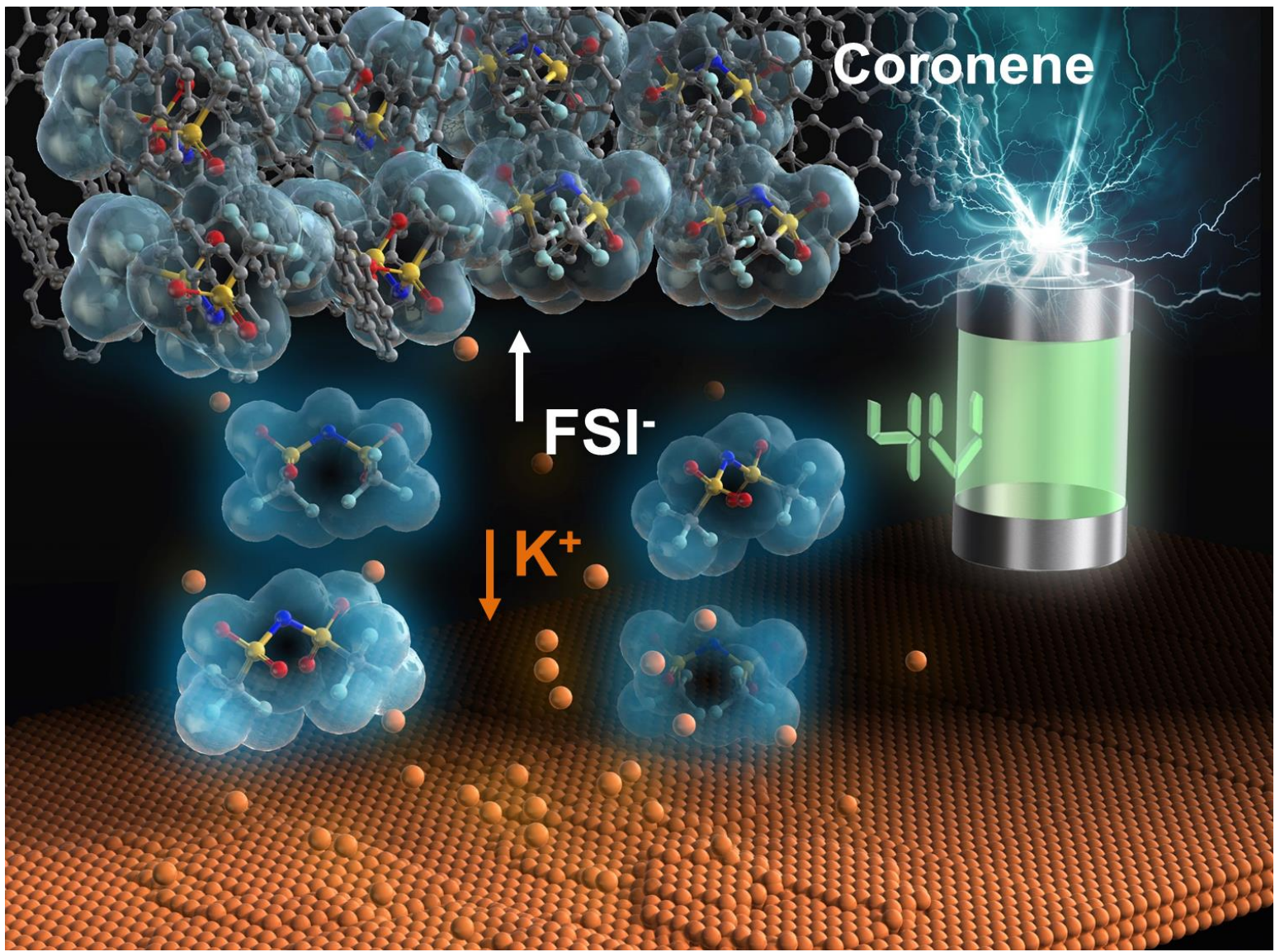

# Coronene: A High-Voltage Anion De-insertion Cathode for Potassium-Ion Battery

Minami Kato[a], Titus Masese[a,b] & Kazuki Yoshii[a]

[a] Research Institute of Electrochemical Energy, National Institute of Advanced Industrial Science and Technology (AIST), 1–8–31 Midorigaoka, Ikeda, Osaka 563–8577, JAPAN

[b] AIST–Kyoto University Chemical Energy Materials Open Innovation Laboratory (ChEM–OIL), Sakyo–ku, Kyoto 606–8501, JAPAN

**Abstract**

Potassium-ion batteries have been envisioned to herald the age of low-cost and high-performance energy storage systems. However, the sparsity of viable components has dampened the progress of these energy devices. Thus, herein, we report coronene, a high-voltage cathode material that manifests a high-voltage of 4.1 V enkindled by anion (de)insertion. This work not only illuminates the broad class of polycyclic aromatic hydrocarbons as prospective cathode materials but also sets a new benchmark for the performance of future organic cathode materials.

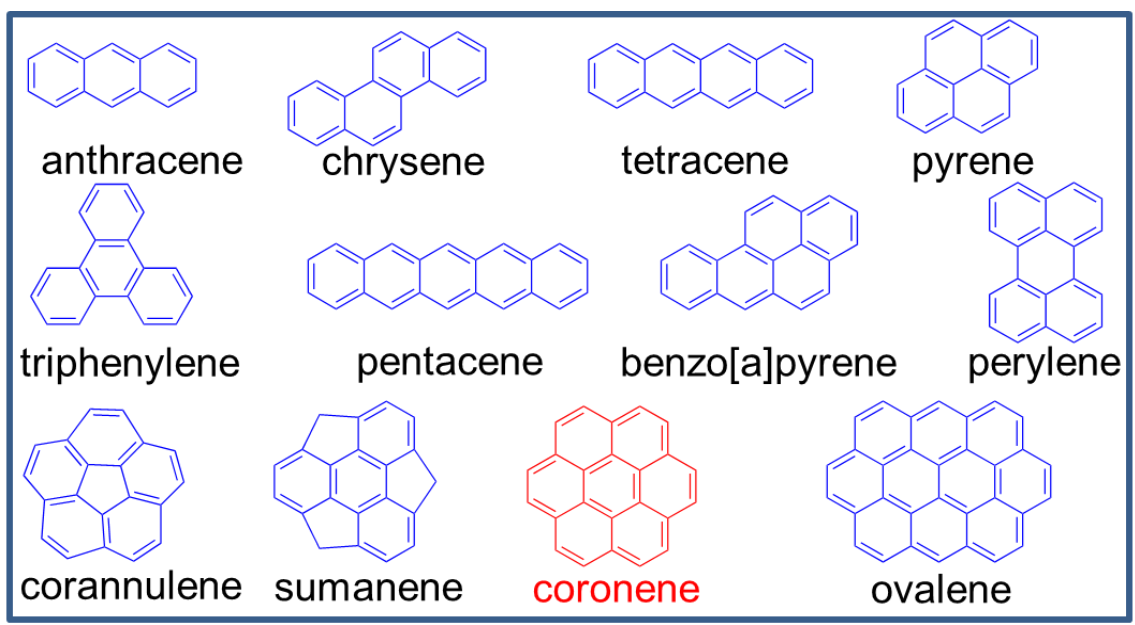

Polycyclic (polynuclear) aromatic hydrocarbons suitable for the development of potassium-ion battery anion de-insertion cathodes. This study was mainly focused on **coronene**, an exemplar of this class of hydrocarbons, that manifests high voltages. The electrochemical properties of other compounds are presented as **Supplementary Information** (available via the link: https://www.rsc.org/suppdata/d1/nj/d1nj00387a/d1nj00387a1.pdf?_ga=2.126917453.1797458518.1634864234-80811797.1634864234).

# 1. <u>INTRODUCTION</u>

The rapidly evolving technological terrain has intensified the necessity for sustainable and high-performance energy storage systems, unveiling potassium as the choice charge carrier for future batteries. Besides having abundant terrestrial reserves, the prominence of potassium-ion as a charge carrier for high-voltage battery systems is primarily fuelled by its superb chemistry as it manifests redox potentials close to lithium and in some cases even lower (in non-aqueous electrolytes).[1] Despite these auspicious prospects, the advancement of potassium-ion batteries is encumbered by the scarcity of viable cathode materials that can facilitate reversible (de) insertion of ions at high voltages.[1,2]

In that context, polyanion-based materials and cyanometallates for instance fluorophosphates (e.g., $KVPO_4F$) and Prussian blue analogues (*e.g.*, $K_2MnFe(CN)_6$) have been pursued as high-voltage cathode contenders, for their abilities to facilitate reversible (de)insertion of potassium ions ($K^+$) voltages around 4 V.[3] Similarly, layered materials (particularly, the honeycomb layered oxides) have also been explored as high-voltage cathode materials, displaying voltages close to 4 V.[3-5] Even so, the sustainability and performance of these inorganic materials are still exiguous. For this reason, organic moieties have been explored for their highly tuneable molecular structures which offer the possibilities of recyclable, flexible and high-performance novel energy devices with more expansive applications.[6, 7]

Amongst the organic moieties pursued, quinones, anhydrides, imides and pyrazinyl compounds have been found to facilitate $K^+$ (de) insertion via *n*-type reactions, whereby the materials accept electrons from electrolytes, concomitantly with the insertion of $K^+$ cations.[8] Although these materials exhibit high $K^+$ (de) insertion capacities, the *n*-type redox reactions in a vast majority of the quinone, anhydride and imide moieties often occur at voltages below 3 V, resulting in potassium-ion batteries with significantly debilitated energy densities. To improve the performance of these organic materials, molecular fine-tuning has been considered as a strategy to raise the output voltages of the *n*-type redox reactions.[8, *ibid*]

Even though typical organic electrode materials facilitate $K^+$ (de)insertion via *n*-type redox reactions, some organic materials have been found to exhibit high voltages by losing electrons (denoted as *p*-type reactions) simultaneously with anion (de)insertion during battery operations. For instance, conducting polymers such as polypyrrole and polyaniline have been found to lose electrons in parallel with the insertion of anions from the electrolyte (*p*-type reaction), delivering higher voltages relative to the organic materials that undergo the *n*-type redox reaction.[9-14] In addition to their high voltage output (in some cases up to 4 V), anion-insertion cathode materials can be coupled with $K^+$-insertion anode materials to assemble full-cells, thereby obviating the need for the costly and sometimes intricate 'pre-potassiation' process. Although literature on *p*-type cathode materials for potassium-ion batteries remains limited, this class of materials has been extensively investigated for lithium- and sodium-ion batteries.[9-14, ibid] In previous studies, radical polymers (such as poly (2,2,6,6-tetramethylpiperidine-*N*-oxyl-4-vinyl ether)) and arylamine polymers (such as poly(triphenylamine)) were reported to facilitate *p*-type reactions, resulting in voltages exceeding 3.5 V.[15,16] As such, the insight into the mechanisms of such materials garnered thus far, will not only foster their adoption into potassium-ion batteries, but also pave way for the materialisation of dual-ion batteries.

Thus, in the quest for sustainable, high-performance potassium battery electrodes, we investigated the broad class of polycyclic compounds which have been known to host *p*-type organic materials. Coronene, an exemplar of the polycyclic aromatic hydrocarbons, emerged not only as an earth-abundant material, but was also evinced to be a high-voltage cathode material, amenable to reversible anion (de)insertion for both lithium- and sodium-ion batteries.[17] Besides its application in cathode materials, this aromatic compound has also been utilised in the preparation of graphite anodes,[18] suggesting significant processing cost reductions can be anticipated in the overall design of future potassium-ion batteries. Furthermore, the topological structure of coronene comprises of benzene rings aligned to each other in a honeycomb configuration (as shown in inset of **Figure 1a**) with close semblance to honeycomb layered oxide materials whose remarkable physicochemical properties have been attributed to their honeycomb alignment.[29] Therefore, herein, we report the electrochemical performance of coronene which exhibits anion (de)insertion at an average voltage of *ca*. 4.1 V in a potassium half-cell.

## 2. RESULTS AND DISCUSSION

The structure morphology of the as-purchased coronene ($C_{24}H_{12}$) was assessed using scanning electron microscopy as shown in **Figure 1a**. The needle-like morphology of coronene crystals is ascertained, with particle sizes in the range of 100 μm. To evaluate the electrochemical performance of coronene, a potassium half-cell configuration utilising bis(fluorosulphonyl)imide (FSI)-based ionic liquid was employed (details furnished in the **METHODS** section). Cyclic voltammograms of coronene composite electrode taken at room temperature reveal conspicuous redox peaks that correspond to anion insertion and de-insertion (extraction) at 4.4 V and 3.8 V, respectively as shown in **Figure 1b**. It should be noted that, the anion being (de)inserted into coronene is $FSI^-$. Redox peaks obtained from subsequent cycling appear to be superimposed, affirming the reversibility of the anion (de)insertion process.

To further assess the voltage characteristics of coronene, galvanostatic (dis)charge measurements were performed on the aforementioned potassium half-cells at varying current densities, as shown in **Figure 2a**. Coronene undergoes a *p*-type reaction delivering a theoretical capacity of about 88 mAh g$^{-1}$; commensurate to a one-electron redox reaction (see **Figure 1b** inset). In the voltage-capacity profiles (**Figure 2a**), coronene displays high voltage plateaux during discharge across the different current

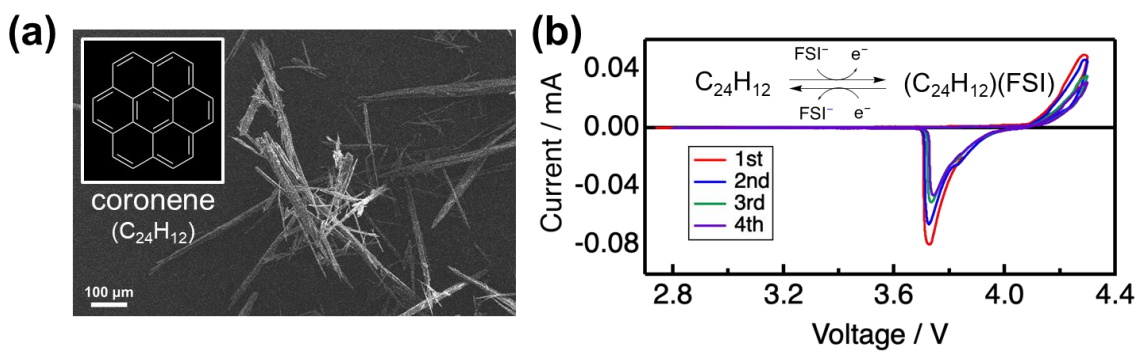

**Figure 1 (a)** Scanning electron microscopy (SEM) images of coronene ($C_{24}H_{12}$) micrometric powders revealing needle-like morphology. **(b)** Cyclic voltammograms of coronene taken at a voltage range of 2.8~4.3 V using potassium bis(fluorosulphonyl)imide (KTFSI)-based ionic liquid. The measurements were done at room temperature and at a scanning rate of 0.1 mV s$^{-1}$.

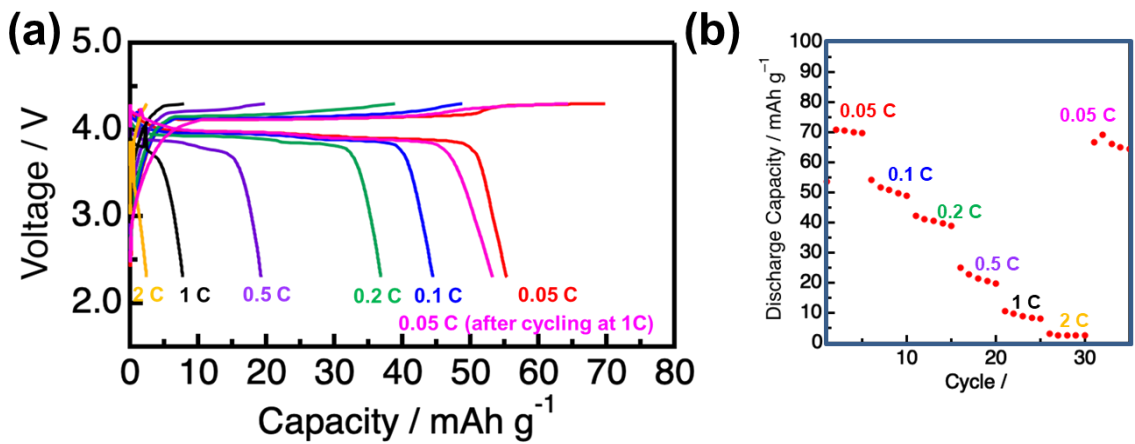

**Figure 2** Electrochemical performance of coronene as a cathode material for potassium batteries. **(a)** Voltage-capacity plots of coronene taken at varying current densities at room temperature. Here 1C corresponds to a current density of 88 mA g$^{-1}$. Galvanostatic (dis)charge measurements were conducted in potassium half-cells using potassium bis(fluorosulphonyl)imide (KFSI)-based ionic liquid electrolyte. **(b)** Rate capability of coronene upon cycling at varying current densities.

densities at room temperature. Voltage plateaux are observed at 4.2 V during charging (FSI$^-$ anion-insertion) and at 4.0 V during discharging (anion extraction), demonstrating coronene to be a 4V-class *p*-type cathode material for potassium-ion batteries. A reversible capacity of approximately 63% was attained at the current density equivalent to the full theoretical (dis)charge of one-electron capacity for 20 h (*viz.*, C/20 rate). Taking into account that no material optimisation strategies such as nanosizing or carbon-coating had been performed on the pristine coronene, the performance observed herein are considered favourable. Although, cycling performed at most current densities (*i.e.*, C/20, C/10 and C/5 rate) yielded fairly satisfactory capacities, the reversible capacity at a high rate of 2C was negligible. Nevertheless, since the initial capacity could be recovered upon cycling at C/20 rate (as further shown in **Figure 2b**), the poor rate performance can be attributed to the slugging anion (de)insertion kinetics.

The electrochemical performance observed here evinces coronene as a high-voltage cathode material capable of reversible anion (de)insertion at an average voltage of 4.1 V. **Figure 3** shows a summary of representative organic cathode materials reported for potassium-ion battery. [15, 19-28] Organic cathode materials that facilitate K$^+$ insertion generally display voltages of below 3 V whereas anion-insertion organic moieties, *vide infra*, exhibit higher voltages (beyond 3 V). The average voltage of 4.1 V reported from coronene (polycyclic aromatic hydrocarbon) reported in this study, is the highest voltage displayed by organic cathode materials for potassium-ion batteries to date. As observed here, anion-insertion organic cathode materials such as coronene generally exhibit low rate capabilities due to the poor anion kinetics along with unsatisfactory cycling performance attributed to the low-molecular-weight of the cathode material which results in high solubility in the electrolyte. Although beyond the scope of the present

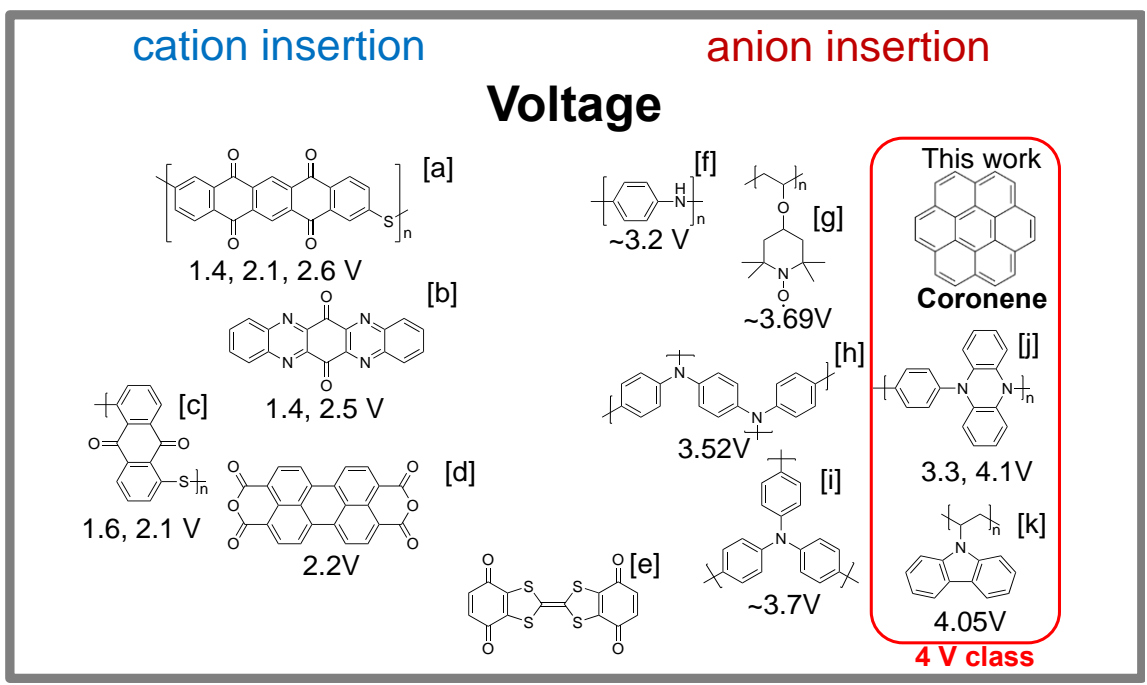

**Figure 3** Voltage reported from anion-insertion cathode materials used in potassium-ion batteries. For clarity, the chemical names of each organic moiety are provided hereafter. [a]: poly(pentacenetetrone sulphide) (**PPTS**), [b]: 5,7,12,14-tetraaza-6,13-pentacenequinone (**TAPQ**), [c]: poly(anthraquinonyl sulphide) (**PAQS**), [d]: 3,4,9,10-perylene-tetracarboxylicacid-dianhydride (**PTCDA**), [e]: (2,3)-(6,7)-bis(1,4-dioxo-1,4-dihydrobenzo)tetrathiafulvalene (**QTTFQ**), [f]: polyaniline (**PANI**), [g]: poly (2,2,6,6-tetramethylpiperidinyloxy methacrylate) (**PTMA**), [h]: poly(*N*,*N*′-diphenyl-*p*-phenylenediamine) (**PDPPD**), [i]: polytriphenylamine (**PTPAn**), [j]: poly(*N*-phenyl-5,10-dihydrophenazine) (**p-DPPZ**) and [k]: poly(N-vinylcarbazole) (**PVK**).[15, 19-28]

study, special synthetic routes (such as *in situ* polymerisation) could be adopted to improve the reaction kinetics (rate capabilities) whilst oligomerisation or polymerisation could be employed to circumvent the solubility of the material (thus, improving the cycling performance). Nevertheless, the high voltage displayed by coronene brings to light the broad family of related polycyclic aromatic hydrocarbons as a test-bench for high-voltage anion (de)insertion materials. Moreover, the abundance of coronene and its application in the scalable synthesis of graphite anode material, ameliorate the feasibility of high-voltage dual ion batteries utilising coronene as the cathode coupled to graphite anode, presenting a sustainable energy storage system solely based on hydrocarbons.

## 4. CONCLUSION

In this study, coronene ($C_{24}H_{12}$), an exemplar polycyclic aromatic hydrocarbon, is assessed as a cathode material for potassium-ion batteries. This organic material is found to facilitate reversible anion-insertion at an average voltage of *ca.* 4.1 V, marking the highest voltage hitherto reported amongst organic cathode materials for potassium-ion batteries. Besides the development of high-voltage energy devices, it is also worthy of mention that coronene is also used in the preparation of graphite anodes, unveiling its potential to actualise low-cost, sustainable dual-ion batteries. This work not only underpins the broad class of related polycyclic aromatic hydrocarbons as a prolific proving ground for high-voltage organic moieties that host *p*-type reactions, but also sets coronene as a performance benchmark for organic materials. Ultimately, the advancement of organic electrodes brings humankind closer to the reality of flexible energy devices that promise to expand the realm of portable electronic technology especially wearable medical devices. [7]

### Declaration of Competing Interests
The authors declare no competing interests.

### Acknowledgements
T. M. gratefully acknowledges Ms. Kumi Shiokawa, Mr. Masahiro Hirata and Ms. Machiko Kakiuchi for their advice and technical help as we conducted the electrochemical measurements. This work was conducted in part under the auspices of the Japan Society for the Promotion of Science (JSPS KAKENHI Grant Numbers 19K15685 and 21K14730), National Institute of Advanced Industrial Science and Technology (AIST) and Japan Prize Foundation.

## Supplementary material

The online link to Supplementary material (which includes the experimental details) associated with this article is available at the following link:
https://www.rsc.org/suppdata/d1/nj/d1nj00387a/d1nj00387a1.pdf?_ga=2.126917453.1797458518.1634864234-80811797.1634864234